\documentclass[aps, prd, 10pt, notitlepage, superscriptaddress, nofootinbib,numbers,showpacs]{revtex4-1}

\usepackage[utf8]{inputenc}
\usepackage[T1]{fontenc}
\usepackage{anyfontsize}
\usepackage{amsmath}
\usepackage{amssymb}
\usepackage{amsfonts}
\usepackage{mathrsfs}
\usepackage{bm} 
\usepackage{microtype}

\usepackage{graphicx}
\usepackage{epsfig}

\usepackage[dvipsnames]{xcolor}
\usepackage[colorlinks=true,linkcolor=blue,citecolor=blue]{hyperref}
\hypersetup{
    colorlinks=true,
    citecolor=Purple,
    linkcolor=Purple,
    urlcolor=Purple,
    linktocpage=true,
    breaklinks=true
}

\usepackage{orcidlink}

\usepackage[capitalize]{cleveref}

\begin{document}
\title{Topological Thermodynamics of Black Holes: Revisiting the Methods of Winding Numbers Calculation}

\author{A. A. M. Silva}
\email{anderson.alves@fisica.ufc.br}
\affiliation{Departamento de F\'isica, Universidade Federal do Cear\'a, Caixa Postal 6030, Campus do Pici, 60455-760 Fortaleza, Cear\'a, Brazil.}

\author{G. Alencar}
\email{geova@fisica.ufc.br}
\affiliation{Departamento de F\'isica, Universidade Federal do Cear\'a, Caixa Postal 6030, Campus do Pici, 60455-760 Fortaleza, Cear\'a, Brazil.}

\author{C. R. Muniz}
\email{celio.muniz@uece.br}
\affiliation{Universidade Estadual do Cear\'a, Faculdade de Educa\c c\~ao, Ci\^encias e Letras de Iguatu, 63500-000, Iguatu, CE, Brazil.}

\author{M. Nilton}
\email{matheus.nilton@fisica.ufc.br}
\affiliation{Universidade Federal Rural do Semi-Árido, 59515-000, Campus Angicos, Angicos, Rio Grande do Norte, Brazil.}

\author{R. R. Landim}
\email{renan@fisica.ufc.br} 
\affiliation{Departamento de F\'isica, Universidade Federal do Cear\'a, Caixa Postal 6030, Campus do Pici, 60455-760 Fortaleza, Cear\'a, Brazil}

\date{today; \LaTeX-ed \today}
\begin{abstract}
In this paper,  the equivalence between two methods for computing winding numbers is established: the approach of $\phi-$mapping topological current and the residue method. The methods are shown to be equivalent when the condition $M''S' - S''M' \neq 0$ holds, while deviations appear when this relation fails, signaling subtle connections between mass $M(r_h)$, entropy $S(r_h)$, and topological structure, with $r_h$ being the horizon radius. We first verify this equivalence to Schwarzschild and Reissner–Nordström black holes, recovering known classifications and confirming the consistency of our approach with respect to the validity of the above condition. We then extend the analysis to four-dimensional black strings, regarded as cylindrically symmetric black hole solutions in asymptotically AdS spacetimes. Our results show that both neutral and charged black strings possess the same global topological number, $W=+1$, implying that electric charge does not influence their topological classification. This insensitivity to charge, mirrors earlier findings for BTZ black holes, suggesting that it may represent a universal property of cylindrically symmetric black holes in AdS backgrounds.

\end{abstract}
\pacs{04.50.Kd,04.70.Bw}
\maketitle
\def\HMS{{\scriptscriptstyle{HMS}}}

\section{Introduction}
\label{S:intro}

The study of black holes has gained renewed momentum in recent years, driven by the advent of high-precision observational techniques. Remarkable results include the direct detection of gravitational waves by LIGO and the first image of a supermassive black hole obtained by the Event Horizon Telescope (EHT) \cite{LIGOScientific:2016aoc,EventHorizonTelescope:2019dse}. Despite these advances, the classical description of black holes remains incomplete, as their interiors are plagued by spacetime singularities. To gain further insight into the nature of such singularities and the thermodynamic and topological properties of horizons, it is often useful to investigate simplified but closely related systems. In this context, black strings provide a useful framework for theoretical investigations. Their extended symmetry makes them more tractable analytically, while still retaining key physical features of black holes.

Recently, the authors in \cite{Wei:2022dzw} suggest that black hole solutions are not only characterized by their thermodynamic quantities, but also by their global topological features in parameter space. These features are interpreted as topological defects in that space, whose classification depends on the winding numbers associated to the thermodynamics potentials. This viewpoint is very useful for distinguishing the different states of the black hole and allows us to identify their universal structures. Since then, several works have further developed this approach \cite{Wei:2024gfz,Wu:2024asq, Gogoi:2025ied, Sekhmani:2025epe, Rani:2025mip,Bao:2025wdp,Nashed:2025njb,Ahmed:2025usm, Chen:2025nto}. Motivated by these developments, we briefly review the thermodynamic topological method in the following.

The starting point for the topological method is based on the concept of off-shell generalized free energy \cite{York:1986it}. In this case, we consider a system consisting of a black hole with mass $M$ and entropy $S$, immersed in a cavity of fixed temperature $1/\tau$. The free energy of this system is described by the expression
\begin{equation}
\mathcal{F}=M-\frac{S}{\tau},
\end{equation}
which naturally extends the standard thermodynamic potential to configurations away from equilibrium. This expression reduces to the on-shell energy of the system when the parameter $\tau$ coincides with the inverse of the Hawking temperature of the black hole, that is, when $1/\tau=T_{H}$. The off-shell construction provides a bridge between the gravitational path integral formulation and the thermodynamic classification of black hole states, making it a powerful tool for topological analyzes \cite{Wei:2022dzw}.

Once the free energy of the system is determined, we can use it together with an auxiliary parameter $\Theta\in(0,\pi)$ to construct a vector field $\vec\phi$ whose components are given by
\begin{equation}
\vec{\phi}=(\phi^{r_h},\phi^{\Theta})=\left(\frac{\partial\mathcal{F}}{\partial r_h},-\frac{\cos{\Theta}}{\sin^2{\Theta}}\right).
\end{equation}

The physically relevant states of the black hole are identified as the zero points of the radial component of $\vec{\phi}$, that is, $\phi^{r_h}=0$. The zero points are identified as topological defects in the space generated by the parameters $(r_h,\Theta)$, whose classification is done using Duan's $\phi-$mapping topological current theory (or just $\phi-$mapping) \cite{Duan:1979ucg}, whereby each zero is endowed with an integer-valued charge, the winding number $w$ \cite{Wei:2022dzw}, that encodes the topological nature of the corresponding black hole solution.

The winding numbers associated with each zero point have a thermodynamic interpretation for black hole states. Locally, a positive winding number ($w=+1$) is characterized by thermodynamically stable states, while a negative winding number ($w=-1$) corresponds to unstable states. When considered globally, the total topological number is obtained by summing all winding numbers, providing a topological classification of black holes that is independent of the details of the system. This dual role of verifying the local stability and global classification of black holes is what makes the topological method a powerful tool for the study of black holes. In particular, most studies have focused on spherically symmetric solutions. However, it is also important to investigate how these topological features change in spacetimes with different symmetries. In this context, black strings provides an ideal model for analyzing how dimensionality and symmetry shape modify the topological classification of black holes.

In this work, we use a method based on the residue theorem to compute winding numbers and applied it to investigate the topological classification of neutral and charged black strings. Our analysis shows that the residue method coincides with the approach of $\phi-$mapping only when the functions $M(r_h)$ and $S(r_h)$ satisfy the relation $M''S'-S''M'\neq0$. Applying this method to black strings, we found that both neutral and charged solutions share the same global topological number, $W=+1$, indicating that the presence of charge does not alter their topological classification. A comparable result was reported in \cite{Chen:2025nto}, where the BTZ black hole was shown to display the same property, suggesting that this may represent a universal feature of cylindrically symmetric black hole solutions in asymptotically AdS spacetimes. The paper is organized as follows: In the Section II we present the $\phi-$mapping and residue methods for computing the winding numbers; the section III we showed the equivalence between two methods and proposed an interpretation for winding number, and we check the equivalence of two methods for Schwarzschild and Reissner-Nordstr\"om black holes.  The section IV presents the application of the topological method to neutral and charged black strings, and Section V summarizes our conclusions.

\section{Topological black holes: $\phi-$mapping and residue methods}
\label{S:Fieldeq}

\begin{figure}[h]
\includegraphics[width=8cm]{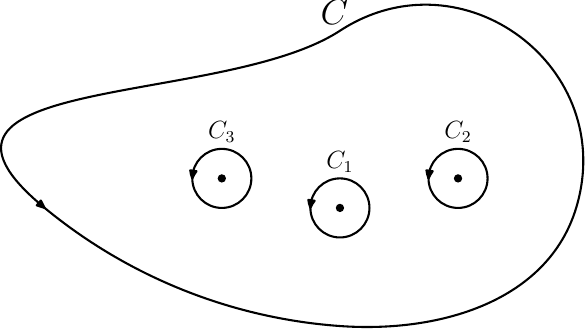}
\caption{The integral over $C$ is equal to the sum over $C_i$}
\label{contornos2-1-mps}
\end{figure}

Recently, \cite{Wei:2022dzw} show that black holes can be viewed as topological defects.  They introduce a generalized free energy
\begin{equation}\label{GFE}
    \mathcal{F}=E-\frac{S}{\tau},
\end{equation}
with $\tau >0$ and a  bidimensional vector
\begin{equation}\label{phi}
    \vec{\phi}=\left(\frac{\partial\mathcal{F}}{\partial r_h},-\cot{\theta}\csc{\theta}\right).
\end{equation}
The zeros points of $\vec\phi$ are in $\theta=\pi/2$ and $\tau=1/T$, where $T$ is the Hawking temperature. Associated to $\vec{\phi}$ we introduce a topological current. 
\begin{equation}\label{j}
    j^\mu=\frac{1}{2\pi}\epsilon^{\mu\nu\rho}\epsilon_{ab}\partial_\nu n^a \partial_\rho n^b, 
\end{equation}
where $\mu,\nu,\rho=0,1,2$, $a,b=1,2$ and
\begin{equation}\label{n}
    n^1=\frac{\phi^1}{\phi}, \quad n^2=\frac{\phi^2}{\phi},\quad \phi=|\vec{\phi}|.
\end{equation}
The current (\ref{j}) is algebraically conserved and  null except at $\phi=0$ \cite{Wei:2022dzw}.
From (\ref{j}), we can write $j^0$ as
\begin{equation}
    j^0=\frac{1}{\pi}(\partial_1 n^1 \partial_2 n^2-\partial_2 n^1 \partial_1 n^2).
\end{equation}
For points where  $\phi\ne0$, with ${\phi^a}$  smooth functions, we have
\begin{equation}
    j^0=\frac{1}{\pi}(\partial_1 Q-\partial_2 P), \quad Q=n^1\partial_2 n^2, P=n^1\partial_1 n^2
\end{equation}
Therefore, for the region $D$ where $\phi$ is not zero, we can apply the Green's theorem 
$$
 0=\int_D j^0 d^2x=\frac{1}{\pi}\int_D (\partial_1 Q-\partial_2 P)d^2x=\frac{1}{\pi}\oint_\Sigma Pdx^1+Qdx^2=\frac{1}{\pi}\oint_\Sigma n^1 dn^2.
$$
For a contour $C$ that involves all zeros of $\phi$, we define the topological number

\begin{equation}\label{W}
   W=\frac{1}{\pi}\oint_C n^1dn^2=\frac{1}{\pi}\sum_{i=1}^N \oint_{c_i}n^1dn^2,
\end{equation}
where $c_i$ are an arbitrary  contour around each zero point of $\phi$ (see Fig. \ref{contornos2-1-mps}). In order to calculate the integral over a circular contour, consider $\vec{\phi}=(f(x),g(y))$, where $f(x_0)=g(y_0)=0$. For convenience, we choose $g'(y_0)=1$, to adapt eq. (\ref{phi}). For a circular contour centered on $(x_0,y_0)$ we parametrize as $(x_0+\varepsilon \cos t,y_0+\varepsilon\sin t)$. Then we have

\begin{equation*}
        f(x(t))=\varepsilon f'(x_0)\cos t+O(\varepsilon),
\end{equation*}
and

\begin{equation*}
    g(y(t))=\varepsilon\sin t+O(\varepsilon).
\end{equation*}
Since the integral contour around $(x_0,y_0)$ is arbitrary,
\begin{equation*}
    \lim_{\varepsilon\rightarrow0}\oint_{c_\varepsilon}n^1dn^2=\int_0^{2\pi}\frac{f'(x_0)^3\cos^2t}{(f'(x_0)^2\cos^2t+\sin^2t)^2}dt=\pi \left(\frac{f'(x_0)}{|f'(x_0)|}\right)^3=\pi\frac{f'(x_0)}{|f'(x_0)|},\quad f'(x_0)\ne0 \quad \mbox{and}\quad 0,\quad f'(x_0)=0.
\end{equation*}
Therefore, the topological number for $\vec{\phi}$  is given by
\begin{equation}\label{W1}
    W=\sum_{i=1}^N \mbox{sgn}\left.\left(\frac{\partial^2\cal{F}}{\partial r_h^2}\right)\right|_{r_h=r_i},
\end{equation}
where $r_i$ are the zeros of $\frac{\partial\mathcal{F}}{\partial r_h}$ and 

\begin{equation}\label{sgn}
\mbox{sgn}(x)=\left\{
\begin{aligned}
~1,&& x>0\\
~0, && x=0\\
-1,&& x<0
\end{aligned}
\right.
\end{equation}

\noindent is the standard sign function, also defined at $x=0$, which allows us to analyze the topological parameters and their relations with thermodynamic variables.

Another method was proposed in \cite{Fang:2022rsb} to calculate the topological number using residues. From the definition of generalized free energy $\partial_{r_h}{\cal{F}}=0$ implies that
\begin{equation}\label{tau}
    \tau=g(r_h).
\end{equation}
We restrict  the solutions of eq. (\ref{tau}) such that $r_h=r_i$ are real.
We define a function in the complex plane by 
\begin{equation}\label{RZ}
    R(z)=\frac{1}{\tau-g(z)},
\end{equation}
which is analytic, except for $z=z_i$, where $\tau-g(z_i)=0$. In the case of $z_i=r_i$, the residues of  $R(r_i)$ are real. The winding number is defined as
\begin{equation}\label{wi}
    w_i=\frac{Res R(r_i)}{|Res R(r_i)|}=\mbox{Sgn}(Res R(r_i)),
\end{equation}
now
\begin{equation}\label{Sgn}
\mbox{Sgn}(x)= \frac{x}{|x|} = \left\{
\begin{aligned}
~1,&& x>0,\\
-1,&& x<0,
\end{aligned}
\right.
\end{equation}
where the $\mbox{Sgn}(x)$, written with an uppercase initial letter, denotes the restricted standard sign function, defined only for nonzero arguments.

\section{The equivalence of two methods, conditions and proposed interpretation for winding number}

We will now show the equivalence of the two methods. From the eq. (\ref{GFE}) we have

\begin{equation}\label{GFE1}
    \frac{\partial{\cal{F}}}{\partial r_h}=M'-\frac{S'}{\tau}=0\rightarrow\tau=\frac{S'}{M'}, \quad M'\ne0,
\end{equation}

\noindent where $M'=dM(r_h)/dr_h$ and $S'=dS(r_h)/dr_h$ at $r_h=r_i$. Therefore, 

\begin{equation}\label{F2}
    \left.\frac{\partial^2{\cal{F}}}{\partial r_h^2}\right|_{r_h=r_i}=M''-S''\frac{M'}{S'}=(M''S'-M'S'')/S'\rightarrow \mbox{sgn}\left.\left(\frac{\partial^2\cal{F}}{\partial r_h^2}\right)\right|_{r_h=r_i}=\mbox{sgn}(M''S'-M'S''), \quad S'>0.
\end{equation}

\noindent Now, from eq. (\ref{RZ}), for $g'(r_i)\ne0$, the residue of $R(r_i)$ is given by

\begin{equation}\label{res1}
    ResR(r_i)=-\frac{1}{g'(r_i)}=\frac{M'^2}{(M''S'-S''M')}\rightarrow \mbox{Sgn}(Res R(r_i))=\mbox{Sgn}(M''S'-S''M').
\end{equation}

\noindent Note that the two procedures are equivalent only if $M''S'-S''M'\ne0.$ In the $\phi-$mapping method when $M''S'-S''M'=0$, this gives $w_i=\mbox{sgn}(0)=0$, but in the residue method $w_i=\mbox{Sgn}(g'''(r_i))$. It is very important to mention that $w_i$ is the sign of specific heat at $r_h=r_i$. To show this, we use the definitions

\begin{equation}\label{T}
    T=\frac{dM}{dS}=\frac{\frac{dM}{dr_h}}{\frac{dS}{dr_h}},
\end{equation}

\noindent and

\begin{equation}\label{C1}
    C=\frac{dM}{dT}=\frac{\frac{dM}{dr_h}}{\frac{dT}{dr_h}}.
\end{equation}\label{C2}

\noindent From the equations above at $r_h=r_i$ we obtain

\begin{equation}\label{sC}
    C=\frac{M'S'^2}{M''S'-M'S''}\rightarrow \mbox{Sgn}(C)=\mbox{Sgn}(M''S'-M'S''),\quad M'>0,
\end{equation}

\noindent  where $C>0$  indicates the local stability for the physical system, and in the topological context, as suggested in \cite{Wei:2022dzw}. It is worth emphasizing that the condition $M''S' - M'S'' = 0$ corresponds to a divergence of the heat capacity and may therefore signal a phase-transition point. However, this local condition should not be confused with the global topological number $W$. A vanishing global number, such as $W=0$, means that the contributions of local winding numbers cancel each other. Thus, $W=0$ reflects the coexistence of branches with opposite local thermodynamic stability, while the divergence of $C$ is associated specifically with the zeros of $M''S' - M'S''$.

\subsection{Applications of the two methods for the Schwarzschild and Reissner-Nordstr\"om black holes.}

\noindent As an application of the $\phi-$mapping method, we consider the Schwarzschild black hole, where $M(r_h)=r_h/2$ and $S(r_h)=\pi r_h^2$. Then
$$
{\cal{F}}=\frac{r_h}{2}-\frac{\pi r_h^2}{\tau}\rightarrow \frac{\partial {\cal{F}}}{\partial r_h}=\frac{1}{2}-\frac{2\pi r_h}{\tau}.
$$
We have one zero for $\partial_{r_h}{\cal{F}}=0$, that is $4\pi r_0=\tau$ and 
$$
\left.\frac{\partial^2\cal{F}}{\partial r_h^2}\right|_{r_h=r_0}=-\frac{2\pi}{\tau}\rightarrow W=w_0=\mbox{sgn}\left(-\frac{2\pi}{\tau}\right)=-1.
$$

Let us now analyze Schwarzschild's case from the point of view of the equivalence relation between the methods, given by equations (\ref{F2}) and (\ref{res1}). We have, for Schwarzschild

\begin{equation}
    M''S' - M'S'' = -\pi \neq 0
\end{equation}

\noindent that is, for any values of $r_i$, both methods are equivalent. Furthermore, we also have at $r_h = r_i$, $w_i = \mbox{Sgn}(M''S' - M'S'') = -1$ and, since $\mbox{Sgn}(C) = w_i$, then $\mbox{Sgn}(C) = -1$, and we conclude that the Schwarzschild  black hole is topologically unstable in $r_h = r_i$. Globally, we have $W = -1$.

For the Reissner-Nordstr\"om (RN) black hole, we have $M(r_h) = \frac{r_h^2 + Q^2}{2r_h}$ and $S(r_h) = \pi r_h^2$ and then

$$
{\cal{F}}=\frac{r_h}{2}+\frac{Q^2}{2r_h}-\frac{\pi r_h^2}{\tau}\rightarrow \frac{\partial {\cal{F}}}{\partial r_h}=\frac{1}{2}-\frac{Q^2}{2r_h^2}-\frac{2\pi r_h}{\tau}, ~\frac{\partial^2{\cal{F}}}{\partial r_h^2}=\frac{Q^2}{r_h^3}-\frac{2\pi}{\tau}
$$
The points where $\partial_{r_h}{\cal{F}}=0$ gives us
$$
\frac{1}{2}-\frac{Q^2}{2r_i^2}=\frac{2\pi r_i}{\tau}.
$$
Since $r_h>0$ and $\tau>0$, we have the conditions $r_i>|Q|$. Like this

$$
\left.\frac{\partial^2\cal{F}}{\partial r_h^2}\right|_{r_h=r_i}=\frac{3Q^2}{2r_i^3}-\frac{1}{2r_i}.
$$
We have only root for $r_i<\sqrt{3}|Q|$ with $w_1=1$ and a second root for  $r_i>\sqrt{3}|Q|$ with $w_2=-1$, which gives the topological charge $W=0$. In the case of two equal roots $r_1=r_2=\sqrt{3}Q$, we also have $W=0$ for the $\phi-$mapping while for residue we have $W=\mbox{Sgn}(g'''(\sqrt{3}|Q|))=-1$.

Analyzing for way of equivalence relations of methods, analogously to Schwarzschild, we have for $r_h=r_i$

\begin{equation}
    M''S' - M'S'' = \frac{\pi}{r_i}(3Q^2 - r_i^2),
\end{equation}

\noindent which brings the equivalence between methods just for $r_i\neq Q\sqrt{3}$, while for the critical value in $r_i = \sqrt{3}Q$, the methods for the obtaining of winding number $w_i$ are not equivalent for the Reissner-Nordstr\"om case, but obtained from different forms as mentioned before. Therefore, considering the regions in which the methods are equivalent, we have:

\begin{itemize}
    \item for $M''S' - M'S''> 0$, consequently $w_i = \mbox{Sgn}(C) = \mbox{Sgn}(M''S' - M'S'') = +1$, we must have $r_i < Q\sqrt{3}$;
    \item for $M''S' - M'S'' < 0$, consequently $w_i = \mbox{Sgn}(C) = \mbox{Sgn}(M''S' - M'S'') = -1$, we must have $r_i > Q\sqrt{3}$.
\end{itemize}

\noindent Thus, as mentioned before, we have, for the RN black hole, $W = -1+1 =0$ for the topological number.

\section{Topological black strings}\label{S:HCS}

In this section, we will classify the global topological properties of black strings using the $\phi-$mapping method and the equivalence of two methods. This is a great opportunity to test this equivalence between $\phi-$mapping and residue methods and determine the topological charge for both. For this, we use the developments of neutral and charged black strings thermodynamic quantities developed in the literature \cite{Lemos_1995,Lemos_1996,lima2022blackstringbouncetraversable,lima2023chargedblackstringbounce}.

Black strings are particular cases of a class of solutions known as black branes. Black branes arise as higher-dimensional generalizations of black holes, obtained by extending the horizon along $p$ translationally invariant spatial directions \cite{Duff:1987cs}. In general, their horizons have topology $S^{D-p-2} \times \mathbb{R}^p$. The case $p=1$ corresponds precisely to black strings, whose horizons take the form $S^{D-3} \times \mathbb{R}$. These objects can thus be regarded as cylindrical analogues of black holes. Unlike their spherically symmetric counterparts, however, black branes (and in particular black strings) exist only in asymptotically anti–de Sitter (AdS) spacetimes \cite{Bellucci:2010gb}. The AdS background is especially relevant in gravitational physics due to the AdS/CFT correspondence \cite{Becker:2006dvp}, which provides a powerful framework to probe the duality between gravity and gauge theories \cite{Witten:1998qj,Witten:1998zw,Henningson:1998gx,Polchinski:1995mt,Polchinski:1994fq}. A four-dimensional black string solution in cylindrical coordinates was first obtained by Lemos \cite{Lemos_1995}. This static solution in an asymptotically anti–de Sitter (AdS) spacetime features a cylindrical horizon with translational symmetry. Lemos also extended this solution to include electric charge \cite{Lemos_1996}, producing charged black strings with modified horizon structure and thermodynamic properties. Since its introduction, this class of solutions has been widely studied \cite{Carvalho:2022eli,Henriquez-Baez:2022ubu,Cunha:2022kep,Muniz:2022otq,Sriling:2021lpr,Awad:2002cz,lima2022blackstringbouncetraversable,lima2023chargedblackstringbounce,Fatima_2012,Nilton:2022jji}, providing valuable insights into the geometry and physics of cylindrical black holes.

In the context of General Relativity, black strings are solutions to Einstein's equations that encompass a cylindrically symmetric black hole. We consider here the static case, that is, independent of time. In this condition, there are three killing vectors, as shown \cite{Lemos_1995}: $\frac{\partial}{\partial z}$, which corresponds to translational symmetry along the axis, $\frac{\partial}{\partial \phi}$, which encloses periodic trajectories around the axis and $\frac{\partial}{\partial t}$, which corresponds to invariance under temporal translations.

For this type of solution, the cosmological constant must be negative. This requirement follows from the definition $\alpha^2=-\Lambda/3>0$, which ensures that $\alpha$ is real only for $\Lambda<0$ \cite{Lemos_1995}. Thus, the spacetime is asymptotically anti--de Sitter, and the negative cosmological constant supports the cylindrical black string geometry considered here. We are considering the coordinates $x^\mu = (t,r,\phi,z)$ which vary according to the intervals: $t\in(-\infty,+\infty)$, $r\in[0,+\infty)$, $\phi \in [0,2\pi]$ and $z\in(-\infty,+\infty)$. As mentioned previously our study of black strings will be limited to neutral and charged black strings. The neutral black string is characterized by the action

\begin{equation}\label{ação black string estática}
	\mathcal{S} = \frac{1}{16\pi}\int d^4x\sqrt{-g}(R - 2\Lambda),
\end{equation}

\noindent The corresponding metric is \cite{lima2022blackstringbouncetraversable}

\begin{equation}\label{métrica da black string}
	ds^2 = -f(r)dt^2 + \frac{dr^2}{f(r)} + r^2d\phi^2 + \alpha^2r^2dz^2, 
\end{equation}

\noindent with $\alpha^2 \equiv -\frac{1}{3}\Lambda > 0$ as mentioned previously, and $f(r) = \alpha^2 r^2 - \frac{b}{\alpha r}$, in which $b$ is a constant that we can relate to mass and we assume it to be positive \cite{Lemos_1995}. Through (\ref{métrica da black string}), taking $g_{00} = f(r) = 0$, we obtain the event horizon \cite{lima2022blackstringbouncetraversable} in

\begin{equation}
	r_h = \frac{b^{1/3}}{\alpha}.
\end{equation}

\noindent From this expression, we obtain the other thermodynamic quantities: temperature, entropy, energy (or their densities in the context of black string). So for these quantities, we have for temperature of the neutral black string.

\begin{equation}\label{th black string}
	T_H = \frac{3\alpha^2r_{h}}{4\pi}.
\end{equation}

\noindent for the mass of neutral black string, in therms of the $r_h$, we see before that the parameter $b$ can be written in therms of the linear mass density \cite{Lemos_1995}. Following again the development of \cite{lima2022blackstringbouncetraversable}, we define $b = 4\mu$, such that:

\begin{equation}\label{massa da black string}
	\mu = \frac{b}{4} = \frac{\alpha^3r^3_h}{4},
\end{equation}

\noindent for the entropy density, using $d S = \frac{d \mu}{T}$:

\begin{equation}
	S = \frac{\alpha}{2}\pi r^2_h.
\end{equation}

\noindent Finally the heat capacity density obtained by $C_V = \frac{d\mu}{dT} = \frac{d\mu}{dr_h}\frac{dr_h}{dT}$, so

\begin{equation}\label{capacidade termoca black string}
	C_V = \frac{3\alpha^2\pi}{4} r^2_h
\end{equation}

The generalized off-shell free energy density for this case is, therefore, given by

\begin{equation}
    \mathcal{F} = \frac{\alpha^3r_h^3}{4} - \frac{\alpha\pi r_h^2}{2\tau}.
\end{equation}

\begin{figure}[!htpb]
    \centering
    \includegraphics[width=0.5\linewidth]{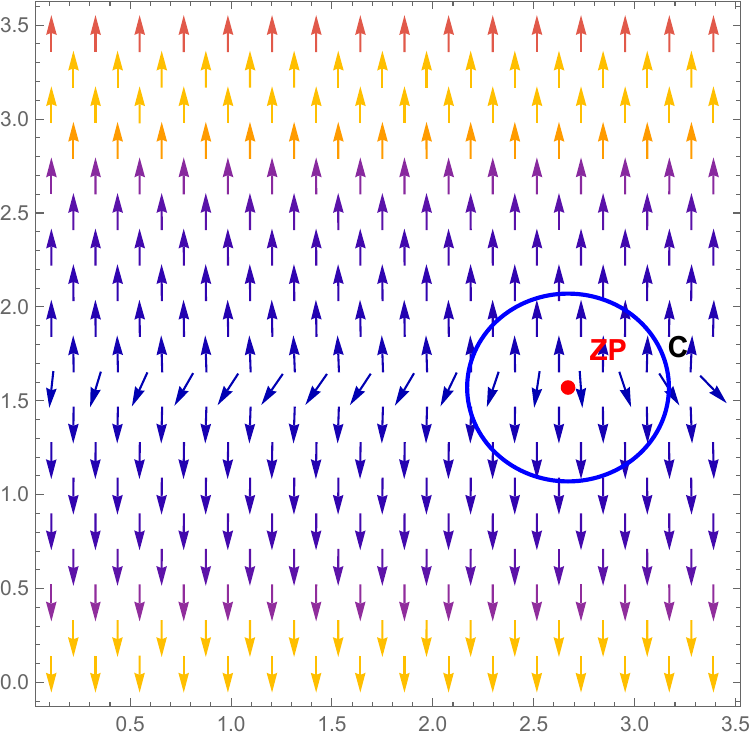}
    \caption{The arrows represent the unit vector field $n$ in a region of the plane $r_h-\Theta$. For this we have considered $\tau=4\pi r_0$ and $\alpha^2r_0=0.005$, where $r_0$ is the radius of the cavity that surrounds the black string. The zero-point(ZP) is marked with a red point and is located in $(r_h/r_0,\Theta)=(2.67,\pi/2)$.}
    \label{fig2}
\end{figure}

\noindent Thus, the components of the vector field are

\begin{eqnarray}
    \phi^{r_h}&=&\frac{\partial \mathcal{F}}{\partial r_h} = \frac{3\alpha^3r_h^2}{4} - \frac{\alpha\pi r_h}{\tau}, \\
    \phi^{\Theta}&=&-\cot{\Theta}\csc{\Theta}.
\end{eqnarray}
The zero-point of the vector field $\phi$ determines the states of the black string, in which they are treated as topological defects in the parameter space $r_h-\Theta$. \noindent For $\left.\frac{\partial \mathcal{F}}{\partial r_h}\right|_{r_h = r_i} = 0$, we have one zero-point, in $r_i = \frac{4\pi}{3\alpha^2\tau}$ and $\Theta=\pi/2$. A plot of the unit vector $n$ in a region of the plane $r_h-\Theta$ is shown in the Fig.\ref{fig2} for the neutral and static black string for $\tau=4\pi r_0$, where $r_0$ is the radius of the cavity. 

We can see in the Fig.\ref{fig2} that there is only one zero-point located in $(r_h,\Theta)=(2.67,\pi/2)$. Therefore, there will be only one winding number representing the local state of the black string. To determine the winding number, we will use equation (\ref{W1}). Thus, we have for a neutral and static black string

\begin{equation}
    \left.\frac{\partial^2 \mathcal{F}}{\partial r_h^2}\right|_{r_h = r_i} = \frac{3\alpha^3r_i}{4}.
\end{equation}

\noindent Since $r_i >0$ and $\alpha > 0$, we have consequently $\left.\frac{\partial^2 \mathcal{F}}{\partial r_h^2}\right|_{r_h = r_i} > 0$ and then, by equation (\ref{W1}), we have $w_1 = +1$ for winding number, which means that the black string has its thermodynamics locally stable. Note that the value of the winding number is independent of the choice of the contour $C$. Since there is only one winding number associated with the zero point, we consequently have that the global topological number $W=+1$. This indicates that the global thermodynamic state of the black string will have essentially the same local state, that is, they both share the same topological number. We can see these features by considering the contour $C$ as the boundary of the entire plane, that is, $(0<\Theta<\pi)\cup(0<r_h<\infty)$. We can see that the vector $\phi$ is pointing downward both in the limit $r_h\rightarrow0$ and $r_h\rightarrow\infty$ with the inclination given by the component $\phi^{\Theta}$ indicating the fact that the global and local topologies of neutral and static black strings are equivalent.

When analyzing the equivalence condition for the $\phi-$mapping and residue methods through equations (\ref{F2}) and (\ref{res1}), where we obtain

\begin{equation}\label{equiv_bs_n}
    M''S' - M'S'' = \frac{3\alpha^4\pi r_i^2}{4} \neq 0, 
\end{equation}

\noindent since $r_i > 0$, which shows that the methods are equivalent in this case.  The winding number is then calculated, so we have:

\begin{equation}\label{winding number bs}
    w_i = \mbox{Sgn}\left(\frac{3\alpha^4\pi r_i^2}{4}\right) = +1.
\end{equation}

Furthermore, by equation (\ref{sC}), since $\mbox{Sgn}(C) = \mbox{Sgn}(M''S' - M'S'') > 0$ for $r_h = r_i$ in the neutral black string, we show the relationship between the winding number and the sign of the heat capacity at $r_h = r_i$ and then conclude that the neutral black string is topologically stable and your topological charge is also $W = +1$.

Now let us consider the case where the black string is charged, in order to analyze how electric charge affects the topological classification of black strings. This case of a black string is characterized by the action

\begin{equation}\label{ação black string estática carregada}
	\mathcal{S} = \frac{1}{16\pi}\int d^4x\sqrt{-g}(R - 2\Lambda - F^{\mu\nu}F_{\mu\nu}).
\end{equation}

\noindent Using the Ansatz with $A_\mu = h(r)\delta^0_\mu$ indicating the electrostatic potential \cite{lima2023chargedblackstringbounce,Lemos_1996}, one obtains
\begin{equation}\label{metric_chatged_bs}
	ds^2 = -f(r)dt^2+\frac{dr^2}{f(r)}+r^2d\phi^2+\alpha^2r^2dz^2,
\end{equation}

\noindent where

\begin{equation}\label{func}
	f(r) = \left(\alpha^2r^2-\frac{b}{\alpha r} + \frac{c^2}{\alpha^2r^2}\right),
\end{equation}

\noindent and $\alpha$ was previously defined, $c^2 = 4\lambda^2$, with $\lambda$ representing the linear charge density of the black string. Here again we have $b = 4\mu$ where $\mu$ represents the mass per unit length of the black string (linear mass density), and the electrostatic potential is $h(r) = -2\lambda/(\alpha r) + h_0$, where $h_0$ is an arbitrary constant \cite{lima2023chargedblackstringbounce}. Using equation (\ref{func}), we obtain the thermodynamic quantities similarly in the neutral black string case. So, for the horizon, we have the approach of \cite{Lemos_1996,lima2023chargedblackstringbounce} and obtain

\begin{equation}\label{horizonte black string carregada}
	r_\pm = \frac{b^{1/3}}{2\alpha}\left(\sqrt{s}\pm\sqrt{2\sqrt{s^2-4q^2}-s}\right),
\end{equation}

\noindent where

\begin{equation}\label{valor de s e q}
	\begin{aligned}
		s &= \left(\frac{1}{2} + \frac{1}{2}\sqrt{1-4\left(\frac{4q^2}{3}\right)^3}\right)^{1/3} + \left(\frac{1}{2} - \frac{1}{2}\sqrt{1-4\left(\frac{4q^2}{3}\right)^3}\right)^{1/3}, \\
		q^2 &= \frac{c^2}{b^{4/3}}.
	\end{aligned}
\end{equation}

\noindent The signs $\pm$ present in (\ref{horizonte black string carregada}) are related to the ``universe'' we are considering. The plus sign refers to the upper universe and the minus sign refers to the lower universe, as in traversable wormholes \cite{Lemos_1996}. For our purpose, we will use $r_+ = r_h$ for the charged black string horizon radius.

For the Hawking Temperature of charged black string $T_{HC}$, we obtain \cite{lima2023chargedblackstringbounce,Fatima_2012}

\begin{equation}\label{Temp_charged_bs}
	T_{HC} = \frac{3\alpha^2r_h}{4\pi} - \frac{c^2}{4\pi\alpha^2r^3_h}.
\end{equation}

For the mass linear density, we have through $\mu = b/4$:

\begin{equation}\label{mass_linear_charged_bs}
	\mu = \frac{1}{4}\left(\alpha^3r^3_h + \frac{c^2}{\alpha r_h}\right).
\end{equation}

\noindent and so, similarly to the neutral black string, we have the entropy density of the charged black string \cite{lima2023chargedblackstringbounce,Fatima_2012}

\begin{equation}\label{entropy_charged_bs}
	S_C = \frac{\pi\alpha r^2_h}{2}.
\end{equation}

With all these ingredients, we can now carry out the study of topological thermodynamics for the charged black string in a way analogous to the neutral black string. In this case the off-shell generalized linear density of free energy is given 

\begin{equation}
    \mathcal{F} = \frac{1}{4}\alpha^3r_h^3 + \frac{c^2}{4\alpha r_h} - \frac{\pi\alpha r_h^2}{2\tau},
\end{equation}

\noindent and therefore we have for the components of the vector field $\phi$:

\begin{equation}
    \begin{aligned}
        \phi^{r_h}&=\frac{\partial\mathcal{F}}{\partial r_h} = \frac{3\alpha^3r_h^2}{4} - \frac{c^2}{4\alpha r_h^2} - \frac{\pi\alpha r_h}{\tau},\\
    \phi^{\Theta}&=-\cot{\Theta}\csc{\Theta}.
    \end{aligned}
\end{equation}

\noindent where for $\left.\frac{\partial \mathcal{F}}{\partial r_h}\right|_{r_h = r_i} = 0$ we have $\frac{1}{\tau} = \frac{3\alpha^2 r_i}{4\pi} - \frac{c^2}{4\pi\alpha^2r_i^3}$ and $\Theta=\pi/2$. Since $\tau>0$, we have the condition $\alpha^2r_i^2>c/\sqrt{3}$. Although the electric charge does not modify the global topological number, it changes the location of the zero point in the parameter space. Indeed, the equilibrium condition depends explicitly on the charge parameter $c$, leading to the constraint $\alpha^2 r_i^2 > c/\sqrt{3}$. Therefore, the charge shifts the thermodynamic stability domain, even though the local winding number remains $w=+1$ and the global classification remains $W=+1$.

\begin{figure}[!htpb]
    \centering
    \includegraphics[width=0.5\linewidth]{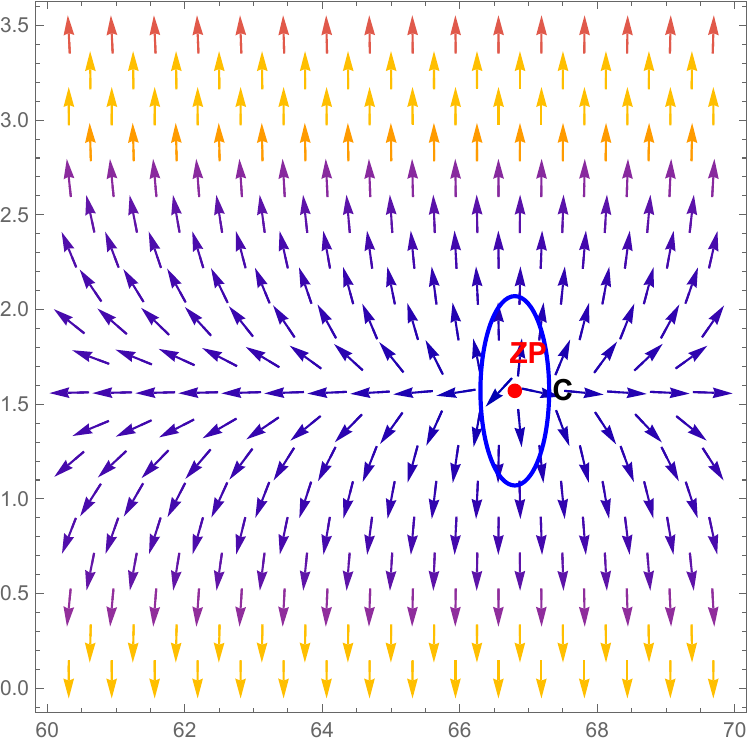}
    \caption{The unit vector field $n$ in a portion of the plane $r_h-\Theta$. We have considered $\alpha^2r_0^2=0.005$, $c^2=0.005$ and $\tau=4\pi r_0$. The zero-point(ZP) is marked with a red point located in $(r_h/r_0,\Theta)=(66.7,\pi/2)$. The blue contour C is enclosing the zero point.}
    \label{fig3}
\end{figure}

A plot of the unit vector $n$ in a region of the plane $r_h-\Theta$ is shown in Fig. \ref{fig3}. As we can see just like the case of the neutral black string, we have only one zero point and therefore must have at least one local state that is thermodynamically stable whose winding number is $w=+1$, which can be easily verified. Indeed, we have

\begin{equation}
    \left.\frac{\partial^2 \mathcal{F}}{\partial r_h^2}\right|_{r_h = r_i} = \frac{3 \left(c^2+\alpha ^4 r_i^4\right)}{4 \alpha  r_i^3}.
\end{equation}

\noindent Since $c^2 > 0$, $\alpha > 0$ and $r_i > 0$, we obtain $\left.\frac{\partial^2 \mathcal{F}}{\partial r_h^2}\right|_{r_h = r_i} > 0$ and then, again by equation (\ref{W1}), we have $w = +1$ for winding number and, consequently we have $W = +1$ for the global topological charge for charged black strings. This happens because the equation $\partial \mathcal{F}/\partial r_h=0$ is strictly positive throughout the domain $r_h>0$, so that the function $\frac{\partial \mathcal{F}}{\partial r_h}(r_h)$ is increasing monotonically and admits only one positive root. Consequently, there is only one zero point in the domain, that is, a single topological defect also for the charged case.

Just like it was done for the neutral black strings and in other examples let's analyze the charged black string under the equivalence of the methods presented before in the equations (\ref{F2}) and (\ref{res1}). We have, for the charged black string in $r_h=r_i$:

\begin{equation}
    M''S'-M'S''= \frac{3 \pi  \left(c^2+\alpha ^4 r_i^4\right)}{4 r_i^2} \neq 0,
\end{equation}

\noindent then, we have $\alpha^2 > 0$, $r_h = r_i >0$ and $c^2>0$, in such a way that the two methods are equivalent whatever $r_i$, and differently of the Reissner-Nordstr\"om case, independent of the charge embedded in $c$. Then we have $M''S'-M'S'' >0$ and consequently, $w_i =\mbox{Sgn}(C)=+1$ like this neutral case, highlighting again the topological stability, but for the case of charged black string.

Just like in the case of the neutral black string the topological classification of the charged black strings according to \cite{Wei:2022dzw} is $W={+1}$ and therefore we can conclude that electric charge does not influence the topological classification of a black string. This behavior has previously been observed in BTZ black holes \cite{Chen:2025nto}, suggesting that such charge insensitivity is not a dimensional peculiarity, but a more general feature of cylindrical symmetry solutions in asymptotically AdS spaces.

In contrast, spherically symmetric and asymptotically flat black holes, such as Schwarzschild and Reissner–Nordström, exhibit an explicit charge dependence on the topological structure of the states. Therefore, the robustness of the topological sorting to electric charge appears to be linked not only to the AdS nature of the background space, but even more profoundly to the cylindrical symmetry of the solution.

\section{Conclusion}\label{S:conclusion}

In this work, we analyze the two methods for computing winding numbers based in $\phi-$mapping and the residue. We showed that the residue method proposed by \cite{Fang:2022rsb}  is equivalent to previously proposed by \cite{Wei:2022dzw}, provided that the condition $M''S' - S''M' \neq 0$ is satisfied. When this condition fails, the residue method yields different results, highlighting additional subtleties in the correspondence between mass, entropy, and topological structure.  

Applying the formalism to the Schwarzschild and Reissner--Nordström cases, we obtained the same results as those found in \cite{Wei:2022dzw}, reinforcing the consistency of our method. We then extended the analysis to black strings, interpreted as four-dimensional solutions with cylindrical symmetry. We found that both the static and charged solutions share the same global classification, $W=+1$, without the appearance of competing branches with opposite winding numbers in the parameter region analyzed here. This indicates that the presence of electric charge does not alter the topological classification of these solutions, in contrast with spherically symmetric and asymptotically flat black holes, where charge plays a fundamental role in determining the topological structure.  

This behavior parallels previous results for BTZ black holes, where it was also shown that electric charge does not affect their topological classification \cite{Chen:2025nto}. Such correspondence suggests that the insensitivity to charge is not accidental, but rather an intrinsic feature of cylindrically symmetric solutions in asymptotically AdS spaces. Thus, our results point to a possible topological universality of black holes with cylindrical symmetry, opening the way for further investigations into the robustness of this property in more general scenarios as, for example, in rotating black strings or in higher dimensions.

\section*{Acknowledgments}
The authors would like to thank Conselho Nacional de Desenvolvimento Cient\'{i}fico e Tecnol\'ogico (CNPq) and Funda\c c\~ao Cearense de Apoio ao Desenvolvimento Cient\'ifico e Tecnol\'ogico (FUNCAP) for the financial support.

\bibliographystyle{apsrev4-1}
\bibliography{ref.bib}
\end{document}